# Bibliometric-enhanced Information Retrieval: 2nd International BIR Workshop

Philipp Mayr*, Ingo Frommholz**, Andrea Scharnhorst, Peter Mutschke*

* GESIS – Leibniz Institute for the Social Sciences, Unter Sachsenhausen 6-8, 50667 Cologne, Germany
philipp.mayr@gesis.org

** Department of Computer Science and Technology, University of Bedfordshire, Luton, UK
ingo.frommholz@beds.ac.uk

**Abstract.** This workshop brings together experts of communities which often have been perceived as different once: bibliometrics / scientometrics / informetrics on the one side and information retrieval on the other. Our motivation as organizers of the workshop started from the observation that main discourses in both fields are different, that communities are only partly overlapping and from the belief that a knowledge transfer would be profitable for both sides. Bibliometric techniques are not yet widely used to enhance retrieval processes in digital libraries, although they offer value-added effects for users. On the other side, more and more information professionals, working in libraries and archives are confronted with applying bibliometric techniques in their services. This way knowledge exchange becomes more urgent. The first workshop set the research agenda, by introducing in each other methods, reporting about current research problems and brainstorming about common interests. This follow-up workshop continues the overall communication, but also puts one problem into the focus. In particular, we will explore how statistical modelling of scholarship can improve retrieval services for specific communities, as well as for large, cross-domain collections like Mendeley or ResearchGate. This second BIR workshop continues to raise awareness of the missing link between Information Retrieval (IR) and bibliometrics and contributes to create a common ground for the incorporation of bibliometric-enhanced services into retrieval at the scholarly search engine interface.

## 1   Introduction

IR and bibliometrics go a long way back. Many pioneers in bibliometrics actually came from the field of IR, which is one of the traditional branches of information science (see e.g. White and McCain, 1998). IR as a technique stays at the beginning of any scientometric[1] exploration, and so IR belongs to the portfolio of skills for any bibliometrician / scientometrician. Used in evaluations, the bibliometric techniques stand and fall with the reliability of identifying sets of work in a field or for an institution. Used in information seeking in large scale of bodies of information, those bibliometric techniques can help to guide the attention of the user to a possible core of information in the wider retrieved body of knowledge.

However, IR and bibliometrics as special scientific fields have also grown apart over the last decades, and with today's 'big data' document collections that bring together aspects of crowdsourcing, recommendation, interactive retrieval and social networks, there is a growing interest to revisit IR and bibliometrics again to provide cutting-edge solutions that help satisfying the complex, diverse and long-term information needs scientific information seekers have. This has been manifesting itself in well-attended combined recent workshops like "Computational Scientometrics" (held at iConference 2013 and CIKM 2013), "Combining Bibliometrics and Information Retrieval"[2] (at the ISSI conference 2013) and last year's ECIR BIR workshop. It became obvious that there is a growing awareness that exploring links between bibliometric techniques and IR is beneficial for both communities (e.g. Wolfram, 2015; Abbasi and Frommholz, 2015). The workshops also made apparent that substantial future work in this direction depends on an ongoing awareness rise in both communities, manifesting itself in concrete experiments/exploration in existing retrieval engines.

There is also a growing importance of combining bibliometrics and information retrieval in real-life applications (see Jack et al., 2014), for instance concerning the monitoring of developments in an area in time. Another example is providing services that support researchers in keeping up-to-date with their field by means of recommendation and interactive search, for instance in 'researcher workbenches' like Mendeley / ResearchGate or search engines like Google Scholar that utilize large bibliometric collections. We hope this workshop will contribute to the identification and further exploration of applications and solutions that bring together both communities. The first bibliometric-enhanced Information Retrieval (BIR) workshop[3] at the ECIR 2014 (Mayr et al., 2014a) has attracted more than 40 participants (mainly from academia) and resulted in three very interactive paper sessions (Mayr et al., 2014b) with lively discussions and future actions. We will build on this experience for the BIR 2015 workshop[4]. Meanwhile a special issue on "Combining Bibliometrics and Information Retrieval" in Scientometrics edited by Philipp Mayr and Andrea Scharnhorst (Mayr and Scharnhorst, 2015) brings together eight papers from experts from bibliometrics / scientometrics / informetrics on the one side and IR on the other.

---

[1] The words bibliometrics, and scientometrics, sometimes even informetrics are used alternatively. While often used interchangeable, scientometrics usually is broader and also includes studies of expenditures, education, institutions, in short all metrics and indicators occurring in quantitative studies of the science system.
[2] http://www.gesis.org/en/events/events-archive/conferences/issiworkshop2013/
[3] http://www.gesis.org/en/events/events-archive/conferences/ecirworkshop2014/
[4] http://www.gesis.org/en/events/events-archive/conferences/ecirworkshop2015/

## 2      Goals, Objectives and Outcomes

Our workshop proposal aims to engage with the IR community about possible links to bibliometrics and complex network theory which also explores networks of scholarly communication. Bibliometric techniques are not yet widely used to enhance retrieval processes in digital libraries, yet they offer value-added effects for users (Mutschke, et al., 2011). To give an example, recent approaches have shown the possibilities of alternative ranking methods based on citation analysis leading to an enhanced IR. Our interests include information retrieval, information seeking, science modelling, network analysis, and digital libraries. The goal is to apply insights from bibliometrics, scientometrics, and informetrics to concrete, practical problems of information retrieval and browsing. More specifically, we ask questions such as:

- How can we build scholarly information systems that explicitly use bibliometric measures at the user interface?
- How can models of science be interrelated with scholarly, task-oriented searching?
- How to combine classical IR (with emphasis on recall and weak associations) with more rigid bibliometric recommendations?
- How to develop evaluation schemes without being caught in too costly setting up large scale experimentation?
- How to combine tools developed in bibliometrics as CitNetExplorer or Science of Science (Sci2) tool with IR?
- And the other way around: Can insights from searching also improve the underlying statistical models themselves?

## 3      Format and Structure of the Workshop

The workshop will start with an inspirational keynote to kick-start thinking and discussion on the workshop topic. This will be followed by paper presentations in a format found to be successful at EuroHCIR 2013 and 2014: each paper is presented as a 10 minute lightning talk and discussed for 20 minutes in groups among the workshop participants followed by 1-minute pitches from each group on the main issues discussed and lessons learned. The workshop will conclude with a round-robin discussion of how to progress in enhancing IR with bibliometric methods.

## 4      Audience

The audiences (or clients) of IR and bibliometrics are different. Traditional IR serves individual information needs, and is – consequently – embedded in libraries, archives and collections alike. Scientometrics, and with it bibliometric techniques, has matured serving science policy. We propose a half-day workshop that should bring together IR and DL researchers with an interest in bibliometric-enhanced approaches. Our interests include information retrieval, information seeking, science modelling, network analysis, and digital libraries. The goal is to apply insights from bibliomet-rics, scientometrics, and informetrics to concrete, practical problems of information retrieval and browsing. The workshop is closely related to the BIR workshop at ECIR 2014 and tries to bring together contributions from core bibliometricians and core IR specialists but having selected those which already operate on the interface between scientometrics and IR. In this workshop we focus more on real experimentation (incl. demos) and industrial participation.

## 5  Output

The papers presented at the BIR workshop 2014 have been published in the online proceedings http://ceur-ws.org/Vol-1143/. Another output of our BIR initiative has been organized after the ISSI 2013 workshop on "Combining Bibliometrics and Information Retrieval" as a special issue in Scientometrics (see Mayr and Scharnhorst, 2015). We aim with the proposed workshop for a similar dissemination strategy, but now oriented towards core IR. This way we build a sequence of explorations, visions, results documented in scholarly discourse, and building up enough material for a sustainable bridge between bibliometrics and IR.